\documentclass{article}
\usepackage{url}
\usepackage{float}
\usepackage{comment}
\usepackage{setspace}
\usepackage{blindtext}
\usepackage{graphicx}
\usepackage{amsmath}
\usepackage{amsfonts}
\usepackage{lipsum}
\usepackage{wrapfig}
\usepackage{multirow}

\begin{document}

\title{\large\bf Android Malicious Application Classification Using Clustering}
\author{\normalsize Hemant Rathore, Sanjay K. Sahay, Palash Chaturvedi and Mohit Sewak \\ {\normalsize BITS, Pilani, Dept. of CS \& IS, Goa Campus, Goa, India} \\ {\footnotesize Email: \{hemantr,ssahay,f20150395,p20150023\}@goa.bits-pilani.ac.in}}

\date{}
	
\maketitle              

\begin{abstract}
	Android malware have been growing at an exponential pace and becomes a serious threat to mobile users. It appears that most of the anti-malware still relies on the signature-based detection system which is generally slow and often not able to detect advanced obfuscated malware. Hence time-to-time various authors have proposed different machine learning solutions to identify sophisticated malware. However, it appears that detection accuracy can be improved by using the clustering method. Therefore in this paper, we propose a novel scalable and effective clustering method to improve the detection accuracy of the malicious android application and obtained a better overall accuracy (98.34\%) by random forest classifier compared to regular method, i.e., taking the data altogether to detect the malware. However, as far as true positive and true negative are concerned, by clustering method, true positive is best obtained by decision tree (97.59\%) and true negative by support vector machine (99.96\%) which is the almost same result obtained by the random forest true positive (97.30\%) and true negative (99.38\%) respectively. The reason that overall accuracy of random forest is high because the true positive of support vector machine and true negative of the decision tree is significantly less than the random forest.
\vspace*{0.1cm}
~
\end{abstract}

\section{Introduction}
The term Malware is derived from \textbf{Mal}icious Soft\textbf{ware}, and initially, malware was developed to show one’s technical skills, but now it has become a profit-driven industry. Over the last decade, android popularity has grown immensely, and its current mobile market share is more than 75\% \cite{marketshare}. The popularity of android Operating System (OS) is because of its open source platform and the availability of a large number of feature-rich applications for it. These applications should ideally be benign, but because of malicious intent of the adversaries, it can be made to perform some unwanted activities in the devices, e.g., stealing the personal information, sending short message service to a premium number, spying on the user, etc.

According to G DATA Security, 3809 new malware samples were detected in 2011 \cite{gdata}. Since then there has been a rapid rise in the number of android malware, and in 2017 more than three million new malware samples have been identified. In this, nearly seven million applications (70\% higher as compared to 2016) were removed from Google Play which was either malware or had some unacceptable content \cite{androidblog}. A report published by McAfee in 2018 shows that there has been an increase of 36\% in AdClick Frauds, 23\% spyware and 12\% Banking Trojan from the previous year \cite{mcafee2018q1}. Another  McAfee Threat Report suggests that by the end of 2017 there were more than twenty million malware samples on the android platform out of which around $\sim$2.5 millions new samples were detected in quarter-3 itself \cite{mcafeedec2017}. For detection of android malware, most of the anti-malware rely on the signature-based detection techniques. But generating and maintaining signature of such a huge number malware is a herculean task. Thus in the last couple of years, researchers have actively started to explore different ways to detect the malware effectively and efficiently. Therefore recently many authors \cite{you2015android} \cite{tam2015copperdroid} \cite{de2015using} \cite{lindorfer2014andrubis} \cite{enck2014taintdroid}  have proposed machine learning as a useful technique to counter the malware, and in this field, the research can be broadly divided into two parts: the feature extraction/selection and methods for the classifications.

\textbf{\textit{Feature extraction}} is an important component of machine learning, and for the effective and efficient classification, features can be extracted without executing the application (static) or during the execution (dynamic) for the detection of malware. Although static analysis gives a better code coverage, it has other limitations. However, with the static feature, several authors have proposed various malware detection models, e.g., Au et al. \cite{au2012pscout} in PScout extracted permissions used in 4 different android OS and found that there are over 75 unique permissions out of which 22\% of the non-system permissions are unnecessary. Lindorfer et al. proposed a fully automated ANDRUBIS system to analyze the android applications from which they found that malicious application on an average request for 13 permission while for good applications the number is just 4.5 \cite{lindorfer2014andrubis}. Ashu et al. used a combination of static opcode frequency and dangerous permissions for malware classification \cite{ashuandroidgroupwise}.  Puerta et al. with the Genome dataset found that opcode frequency is better feature vector than the permissions \cite{de2015using}. 

Malware can also be analyzed while executing them in a controlled environment to find the features, e.g., network traffic, application programming interface, system calls, information flow for the classifications. In this, Enck et al. built TaintDroid to track the runtime information flow and found many instances of private data being misuse in different applications \cite{enck2014taintdroid}.  Tam et al. developed CopperDroid to model well know process-OS interaction and also inter-process and intra-process communications for effective malware detection \cite{tam2015copperdroid}. You et al. in 2015 did a comprehensive analysis using Dalvik opcodes as features to find the potential threats \cite{you2015android}.

Finally, the features obtained by the static or dynamic approach is an important ingredient for the \textbf{\textit{classification}}. In this, Puerta et al. proposed a single class classifier for malware detection based on opcode occurrence and achieved an accuracy of 85\% \cite{de2015using}. Ashu et al. used opcode occurrence on five different classifiers and obtained the highest accuracy of 79.27\% with the functional tree \cite{sharma2018investigation}. Feizollah et al. proposed AndroDialysis, which used intent and permission separately and then combine them to make an extensive feature vector \cite{feizollah2017androdialysis}. Their analysis shows that independently intent and permission achieved a detection rate of 91\% and 83\% respectively. However, combing both the features shows an increase in the detection rate (95.5\%). Wu et al. proposed DroidMat, which collects permission, intent, API Call and applied various machine learning algorithms like k-nearest neighbor and naive bayes for malware detection, and achieved the highest accuracy of 97.87\% \cite{wu2012droidmat}. Arp et al. used permission, intent, API call, network address in the Drebin dataset of more than 5000 malware samples and utilized support vector machine for the classification, and achieved an accuracy of 94\% \cite{arp2014drebin}.

Recently Ashu et al. used opcode frequency as a feature and grouped them based on permission to achieve a detection accuracy of 97.15\% \cite{ashuandroidgroupwise}. Li et al. used permission as the feature vector and used three level pruning for identification of significant permission for effective detection of malware and benign \cite{li2018significant}. They have used 22-most significant permissions and achieved a detection rate of 93.62\%. Rana et al. use different tree based classifiers like a decision tree, random forest, gradient boosting and extremely randomized tree with n-gram approach and shown that with 3-gram and random forest classifier one can achieve an accuracy up to 97.24\% \cite{rana2018evaluation}. Chen et al. used n-gram opcode sequence with exemplar feature selection method and random forest classifier to detect the malware only with 95.6\% correctly with 4-gram approach \cite{chen2018tinydroid}.

The above-proposed methods using different machine learning seem to be not sufficient to identify sophisticated advanced malware. However, it appears that detection accuracy can be improved by using the clustering method. Therefore in this paper, we propose a novel scalable and effective clustering method to improve the detection accuracy of the android malware. Hence, the rest of the paper is organized as follows: Section 2 introduces the dataset and feature extraction. Section 3 contains the experimental analysis and results. Finally, Section 4 concludes the paper.

\section{Dataset and Feature Extraction}
In the area of machine learning, it is very important that how good one can make/train a model to identify the target in new/test data, and in turn, the quality of the trained model depends on the dataset and how the features are extracted/selected from it. Understanding the fact we used Drebin dataset \cite{arp2014drebin} which is one of the largest benchmark malware samples, which consists of 5550 malware samples from more than 20 different families. It also includes all the malware samples from the Android Malware Genome Project \cite{jiang2012dissecting}.

For the benign file, we have collected 8500 android applications between 2016-17 from various sources (Google Play \cite{googleplay}, Third-party app stores, alternate marketplaces, etc.). To test the downloaded files is benign or not, we verified them using VirusTotal \cite{virustotal} (it is a subsidy of Alphabet which is an aggregator of 40-60 antivirus including AVG, Bitdefender, F-Secure, Kaspersky Lab, McAfee, Norton, Panda Security, etc. and provide various API’s to check whether the application is malware/benign) services, and we declare an application as malicious if one or more antivirus from virustotal.com categories it as malware. Thus after verifying all the downloaded samples from VirusTotal, we were left with 5720 samples for the experimental analysis.

For the analysis, we disassemble all the sample file by APKtool \cite{apktool} to extract all the opcodes for the clustering and classifications. For this purpose, we have taken the frequency of the opcode as a feature vector and generated the feature vector for the complete dataset as $11266 \times 256$ matrix, where rows represent different files, and a column represents the frequency of the opcode in that particular file.

\section{Experimental Analysis and Result}
\begin{figure}[h]
	\centering
	\includegraphics[scale=.45]{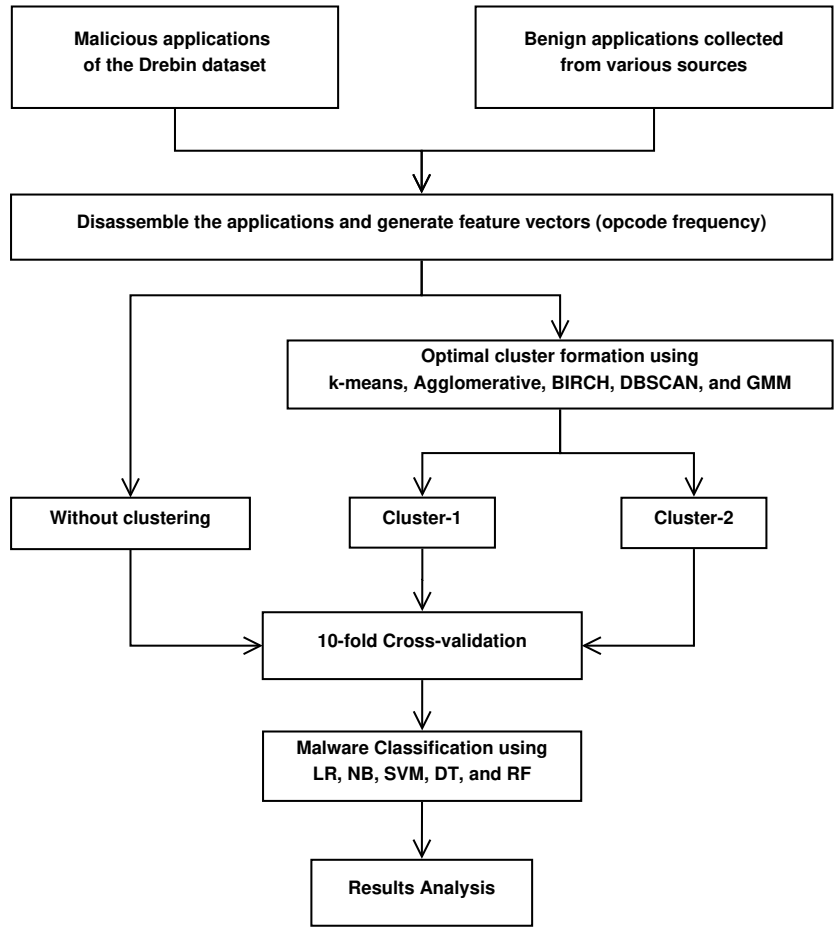}
	\caption{Flowchart of our novel method to improved malware detection. accuracy}
	\label{Flowchart}
\end{figure}

A schematic of the experimental analysis of our novel proposed approach/method is depicted in figure \ref{Flowchart}. We first experimented by the regular method given by the various authors \cite{jiang2012dissecting} \cite{arp2014drebin} \cite{wu2012droidmat} \cite{feizollah2017androdialysis} \cite{de2015using}, i.e., without clustering/grouping the data for comparison of accuracy with our novel method (first we find the optimal number of cluster and then identify the malicious android applications), and we used the scikit-learn \cite{sklean} machine learning library for all the clustering and classifications. Also Numpy python library has been used to handle large multidimensional arrays and matrices \cite{numpy}.

\subsection{Malware Detection without clustering}
To understand the performance of the different type of classifiers by regular used method (i.e., taking the whole data altogether), we used Logistic Regression (LR), Naive Bayes (NB), Support Vector Machine (SVM), Decision Tree (DT), and Random Forest (RF). For the classification, we used all 256 opcodes as features and ten-fold cross-validation (it divides the data into ten equal parts out of which nine parts are used for training the model, and one part is used for testing. This exercise is completed ten times with different combinations of training and testing set and at last average result is taken into account) and the result obtained is shown in figure \ref{crossval}. From the analysis, we found that after three-fold cross-validation the accuracy of classifiers is more or less same. The detailed results of the experiment conducted by all the five classifiers in terms of Accuracy, Recall, and Specificity are given in the table \ref{table_without_cluster}, where

$$	\text{Accuracy}  =  \frac{\text{Number of Malware and Benign correctly classified}}{\text{Total Number of files in the dataset}}$$

$$ \text{True Postive Rate (TPR)} = \frac{\text{Number of Malware correctly classified}}{\text{Number of Malware in the dataset}}$$

$$ \text{True Negative Rate (TNR)} = \frac{\text{Number of Benign correctly classified}}{\text{Number of Benign in the dataset}}$$

$$$$

The analysis also shows that the tree-based RF classifier outperformed the other classifiers with the best accuracy of 95.82\%.

%
\begin{table}[h]
	\begin{center}
		\caption{Accruacy, TPR and TNR obtained without clustering the data.}
		
		\medskip
		
		\label{table_without_cluster}
		\begin{tabular}{|c|c|c|c|c|c|}
			\hline
			\textbf{} & \textbf{\begin{tabular}[c]{@{}c@{}}Logistic\\ Regression\end{tabular}} & \textbf{\begin{tabular}[c]{@{}c@{}}Naive \\   Bayes\end{tabular}} & \textbf{\begin{tabular}[c]{@{}c@{}}Support Vector\\ Machines\end{tabular}} & \textbf{\begin{tabular}[c]{@{}c@{}}Decision \\ Trees\end{tabular}} & \textbf{\begin{tabular}[c]{@{}c@{}}Random  \\ Forest\end{tabular}} \\ \hline
			\textbf{Accuracy} & 87.96 & 76.66 & 84.68 & 93.70 & 95.82 \\ \hline
			\textbf{Recall/TPR} & 92.44 & 94.89 & 69.38 & 95.27 & 95.40 \\ \hline
			\textbf{Specificity/TNR} & 84.16 & 58.49 & 99.96 & 92.65 & 95.73 \\ \hline
		\end{tabular}
	\end{center}
\end{table}


%
\subsection{Clustering based Malware Detection}
\begin{figure}[h]
	\begin{minipage}[b]{0.5\linewidth}
		\includegraphics[trim=0cm 0cm 3.20cm 3.0cm, clip, height=5.2cm, width=6cm]{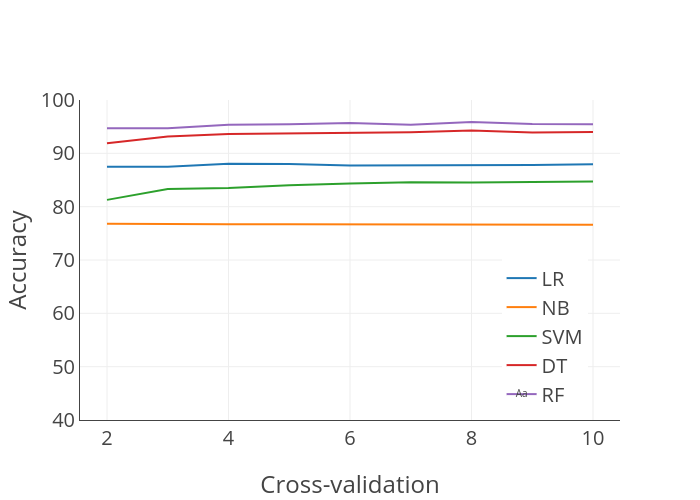}
		\caption{Classifier accuracy with the k-fold Cross-Validation.}
		\label{crossval}
	\end{minipage}
	\begin{minipage}[b]{0.5\linewidth}
		\includegraphics[trim=0cm 0cm 3.50cm 2.30cm, clip, height=5.2cm, width=6cm]{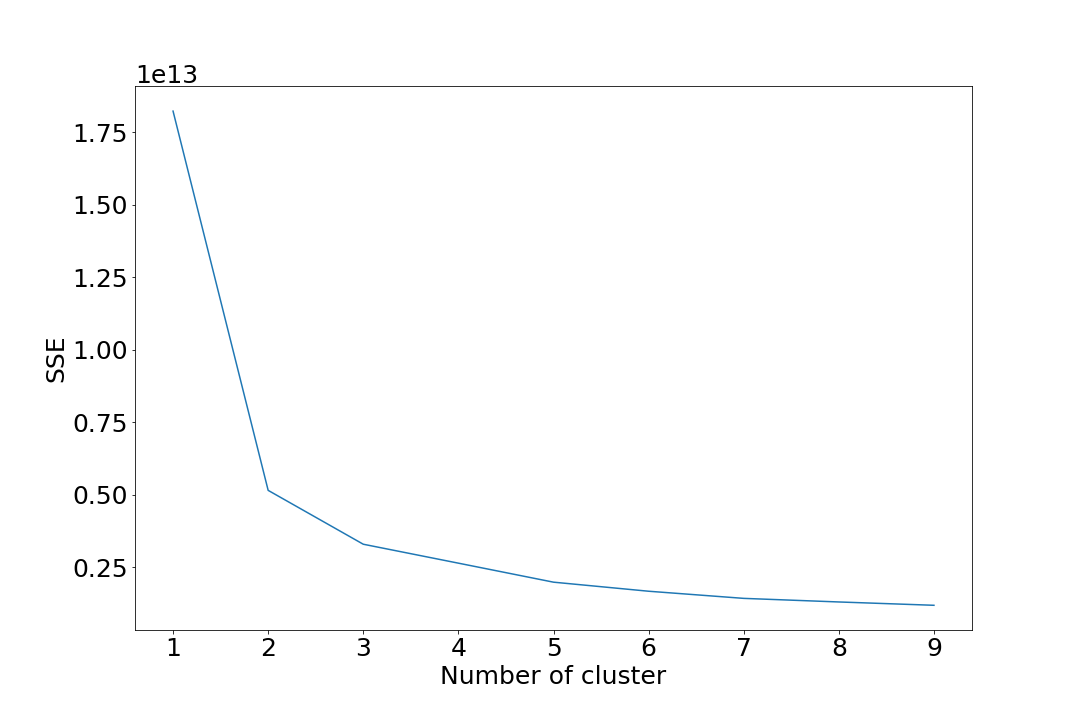}
		\caption{Number of clusters with SSE by k-means algorithm.}
		\label{elbow}
	\end{minipage}
\end{figure}

To improve the detection accuracy of malicious applications, we first studied the five different clustering algorithms (k-means, Agglomerative, BIRCH, DBSCAN, and GMM) to find the clusters in the dataset. The analysis shows (table \ref{ClusterAnalysis}) that the k-means algorithm forms the best cluster (Calinski-Harabasz \cite{calinski1974dendrite} and Silhouette score \cite{rousseeuw1987silhouettes} of the k-means is best among the selected five clustering algorithm). Further, we observed that the two clusters formed by the k-means would be best for the classification of malicious applications (fig \ref{elbow}) and in the obtained cluster, we find that cluster-1 contains total 8760 files and is dominated by malware files whereas the cluster-2 contains total 2780 files dominated by benign files. However, for the analysis, we have balanced the dataset using SMOTE \cite{chawla2002smote}. Now for the classification, we use the same set of classifiers that are used for the classification without clustering and found that the overall accuracy, TPR and TNR obtained by clustering the data is significantly more than without clustering the data and the results obtained are given in table \ref{table_with_clustering}.



\begin{table}
	\begin{center}
	\caption{Number of clusters with different clustering algorithm and its Calinski Harabaz and Silhouette Score.}
	
	\bigskip
	
	\label{ClusterAnalysis}
	\begin{tabular}{|c|c|c|c|}
		\hline
		&&&\\
		& No of Clusters & Calinski Harabaz Score & Silhouette Score \\ 
		&&&\\\hline
		\multirow{4}{*}{\textbf{k-means Clustering}} & \textbf{2} & \textbf{28568.28} & \textbf{0.7377} \\ \cline{2-4} 
		& 3 & 25493.58 & 0.7005 \\ \cline{2-4} 
		& 4 & 22109.38 & 0.6167 \\ \cline{2-4} 
		& 5 & 22971.34 & 0.6013 \\ \hline
		\multirow{4}{*}{\textbf{Agglomerative Clustering}} & 2 & 24706.58 & 0.7256 \\ \cline{2-4} 
		& 3 & 21245.43 & 0.7069 \\ \cline{2-4} 
		& 4 & 20046.19 & 0.6043 \\ \cline{2-4} 
		& 5 & 20939.81 & 0.6052 \\ \hline
		\multirow{4}{*}{\textbf{BIRCH Clustering}} & 2 & 24735.95 & 0.7258 \\ \cline{2-4} 
		& 3 & 20844.19 & 0.7080 \\ \cline{2-4} 
		& 4 & 18494.73 & 0.6656 \\ \cline{2-4} 
		& 5 & 21261.10 & 0.6056 \\ \hline
		\multirow{4}{*}{\begin{tabular}[c]{@{}c@{}}\textbf{Gaussian Mixture Model}\\ \textbf{Clustering}\end{tabular}} & 2 & 8398.36 & 0.4716 \\ \cline{2-4} 
		& 3 & 2738.43 & 0.1967 \\ \cline{2-4} 
		& 4 & 1640.20 & 0.1010 \\ \cline{2-4} 
		& 5 & 4813.65 & 0.2519 \\ \hline
		\multirow{4}{*}{\textbf{DBSCAN Clustering}} & eps = 5000 & 1380.75 & 0.1539 \\ \cline{2-4} 
		& eps = 10000 & 3469.39 & 0.5316 \\ \cline{2-4} 
		& eps = 15000 & 3743.52 & 0.5349 \\ \cline{2-4} 
		& eps = 20000 & 4858.33 & 0.6753 \\ \hline
	\end{tabular}
	\end{center}

\vspace{1.0cm}

	\begin{center}
{\footnotesize
		\caption{Accuracy, TPR and TNR after clustering the data.}
		
		
		\label{table_with_clustering}
		\begin{tabular}{|c|c|c|c|c|c|}
			\hline
			\textbf{} & \textbf{\begin{tabular}[c]{@{}c@{}}Logistic\\ Regression\end{tabular}} & \textbf{\begin{tabular}[c]{@{}c@{}}Naive \\   Bayes\end{tabular}} & \textbf{\begin{tabular}[c]{@{}c@{}}Support Vector\\ Machines\end{tabular}} & \textbf{\begin{tabular}[c]{@{}c@{}}Decision \\ Trees\end{tabular}} & \textbf{\begin{tabular}[c]{@{}c@{}}Random  \\ Forest\end{tabular}} \\ \hline
			\textbf{Accuracy} & 92.23 & 81.01 & 92.29 & 97.92 & 98.34 \\ \hline
			\textbf{Recall/TPR} & 91.49 & 96.46 & 84.59 & 97.59 & 97.30	 \\ \hline
			\textbf{Specificity/TNR} & 92.98 & 65.57 & 99.96 & 98.25 & 99.38 \\\hline
			
		\end{tabular}
}
	\end{center}
\end{table}
\section{Conclusion}
To detect the malware, generally signature-based anti-malware are used which is not good enough to detect the advanced obfuscated malware. Hence time-to-time to detect the advanced malware different machine learning solutions are proposed by various authors. In this, we proposed a novel scalable and effective clustering based android malware detection system. The analysis shows an overall improvement of accuracy with RF (98.34\%) by our proposed clustering method. The accuracy achieved by our approach outperformed the recent accuracy obtained by the authors viz. Ashu et al. (97.15\%), Li et al. (93.62\%) and Rana et al. (97.24\%). Also, the experimental analysis shows that despite the TNR of SVM and TPR of DT are marginally better than the RF, the overall accuracy of RF is best among the tested classifiers, this is because the TPR of SVM and TNR of DT is far below the RF classifier. As the results are significant, therefore we are developing an API for the identification of the android malicious apps, which we will be free to the research community.
\bibliographystyle{plain}
\bibliography{AndroidISDA}

\begin{thebibliography}{10}

\bibitem{gdata}
{G DATA M}obile {I}nternet {S}ecurity.
\newblock Technical report, G DATA, 2017 (Date last accessed 02-Oct-2018).

\bibitem{marketshare}
{S}martphone {OS} {M}arket {S}hare.
\newblock Technical report, ITC, 2017 (Date last accessed 02-Oct-2018).

\bibitem{apktool}
{APKTOOL}.
\newblock Technical report, Apache, 2018 (Date last accessed 02-Oct-2018).

\bibitem{googleplay}
{G}oogle {P}lay.
\newblock Technical report, Google, 2018 (Date last accessed 02-Oct-2018).

\bibitem{androidblog}
{H}ow we fought bad apps and malicious developers in 2017.
\newblock Technical report, Android Developers Blog, 2018 (Date last accessed
  02-Oct-2018).

\bibitem{mcafeedec2017}
{M}c{A}fee {M}obile {T}hreat {R}eport {D}ecember 2017.
\newblock Technical report, McAfee, 2018 (Date last accessed 02-Oct-2018).

\bibitem{mcafee2018q1}
{M}c{A}fee {M}obile {T}hreat {R}eport {Q}1, 2018.
\newblock Technical report, McAfee, 2018 (Date last accessed 02-Oct-2018).

\bibitem{numpy}
{N}um{P}y.
\newblock Technical report, 2018 (Date last accessed 02-Oct-2018).

\bibitem{sklean}
scikit-learn.
\newblock Technical report, 2018 (Date last accessed 02-Oct-2018).

\bibitem{virustotal}
{V}irus{T}otal.
\newblock Technical report, Google, 2018 (Date last accessed 02-Oct-2018).

\bibitem{arp2014drebin}
Daniel Arp, Michael Spreitzenbarth, Malte Hubner, Hugo Gascon, Konrad Rieck,
  and CERT Siemens.
\newblock Drebin: Effective and explainable detection of android malware in
  your pocket.
\newblock In {\em Ndss}, volume~14, pages 23--26, 2014.

\bibitem{au2012pscout}
Kathy Wain~Yee Au, Yi~Fan Zhou, Zhen Huang, and David Lie.
\newblock Pscout: analyzing the android permission specification.
\newblock In {\em Proceedings of the 2012 ACM conference on Computer and
  communications security}, pages 217--228. ACM, 2012.

\bibitem{calinski1974dendrite}
Tadeusz Cali{\'n}ski and Jerzy Harabasz.
\newblock A dendrite method for cluster analysis.
\newblock {\em Communications in Statistics-theory and Methods}, 3(1):1--27,
  1974.

\bibitem{chawla2002smote}
Nitesh~V Chawla, Kevin~W Bowyer, Lawrence~O Hall, and W~Philip Kegelmeyer.
\newblock Smote: synthetic minority over-sampling technique.
\newblock {\em Journal of artificial intelligence research}, 16:321--357, 2002.

\bibitem{chen2018tinydroid}
Tieming Chen, Qingyu Mao, Yimin Yang, Mingqi Lv, and Jianming Zhu.
\newblock Tinydroid: A lightweight and efficient model for android malware
  detection and classification.
\newblock {\em Mobile Information Systems}, 2018, 2018.

\bibitem{de2015using}
Jos{\'e}~Gaviria de~la Puerta, Borja Sanz, Igor Santos, and Pablo~Garc{\'\i}a
  Bringas.
\newblock Using dalvik opcodes for malware detection on android.
\newblock In {\em International Conference on Hybrid Artificial Intelligence
  Systems}, pages 416--426. Springer, 2015.

\bibitem{enck2014taintdroid}
William Enck, Peter Gilbert, Seungyeop Han, Vasant Tendulkar, Byung-Gon Chun,
  Landon~P Cox, Jaeyeon Jung, Patrick McDaniel, and Anmol~N Sheth.
\newblock Taintdroid: an information-flow tracking system for realtime privacy
  monitoring on smartphones.
\newblock {\em ACM Transactions on Computer Systems (TOCS)}, 32(2):5, 2014.

\bibitem{feizollah2017androdialysis}
Ali Feizollah, Nor~Badrul Anuar, Rosli Salleh, Guillermo Suarez-Tangil, and
  Steven Furnell.
\newblock Androdialysis: Analysis of android intent effectiveness in malware
  detection.
\newblock {\em computers \& security}, 65:121--134, 2017.

\bibitem{jiang2012dissecting}
Xuxian Jiang and Yajin Zhou.
\newblock Dissecting android malware: Characterization and evolution.
\newblock In {\em 2012 IEEE Symposium on Security and Privacy}, pages 95--109.
  IEEE, 2012.

\bibitem{li2018significant}
Jin Li, Lichao Sun, Qiben Yan, Zhiqiang Li, Witawas Srisa-an, and Heng Ye.
\newblock Significant permission identification for machine learning based
  android malware detection.
\newblock {\em IEEE Transactions on Industrial Informatics}, 2018.

\bibitem{lindorfer2014andrubis}
Martina Lindorfer, Matthias Neugschwandtner, Lukas Weichselbaum, Yanick
  Fratantonio, Victor Van Der~Veen, and Christian Platzer.
\newblock Andrubis--1,000,000 apps later: A view on current android malware
  behaviors.
\newblock In {\em Building Analysis Datasets and Gathering Experience Returns
  for Security (BADGERS), 2014 Third International Workshop on}, pages 3--17.
  IEEE, 2014.

\bibitem{rana2018evaluation}
Md~Shohel Rana, Sheikh Shah Mohammad~Motiur Rahman, and Andrew~H Sung.
\newblock Evaluation of tree based machine learning classifiers for android
  malware detection.
\newblock In {\em International Conference on Computational Collective
  Intelligence}, pages 377--385. Springer, 2018.

\bibitem{rousseeuw1987silhouettes}
Peter~J Rousseeuw.
\newblock Silhouettes: a graphical aid to the interpretation and validation of
  cluster analysis.
\newblock {\em Journal of computational and applied mathematics}, 20:53--65,
  1987.

\bibitem{sharma2018investigation}
Ashu Sharma and Sanjay~Kumar Sahay.
\newblock An investigation of the classifiers to detect android malicious apps.
\newblock In {\em Information and Communication Technology}, pages 207--217.
  Springer, 2018.

\bibitem{ashuandroidgroupwise}
Ashu Sharma and S.K. Sahay.
\newblock Group-wise classification approach to improve android malicious apps
  detection accuracy.
\newblock In {\em International Journal of Network Security, 2018}, page
  Accepted, 2018.

\bibitem{tam2015copperdroid}
Kimberly Tam, Salahuddin~J Khan, Aristide Fattori, and Lorenzo Cavallaro.
\newblock Copperdroid: Automatic reconstruction of android malware behaviors.
\newblock In {\em NDSS}, 2015.

\bibitem{wu2012droidmat}
Dong-Jie Wu, Ching-Hao Mao, Te-En Wei, Hahn-Ming Lee, and Kuo-Ping Wu.
\newblock Droidmat: Android malware detection through manifest and api calls
  tracing.
\newblock In {\em Information Security (Asia JCIS), 2012 Seventh Asia Joint
  Conference on}, pages 62--69. IEEE, 2012.

\bibitem{you2015android}
Wei You, Bin Liang, Jingzhe Li, Wenchang Shi, and Xiangyu Zhang.
\newblock Android implicit information flow demystified.
\newblock In {\em Proceedings of the 10th ACM Symposium on Information,
  Computer and Communications Security}, pages 585--590. ACM, 2015.

\end{thebibliography}

\end{document}